**E.M. Pokrovskaya**
**PhD in Philosophy, Head of the foreign language department,**
**PhD in Philosophy, Associate professor,**
**"Tomsk State University of Control Systems and Radioelectronics"**
**Tomsk**
**Margarita Y. Raitina**
**PhD in Philosophy, associate professor of the philosophy and sociology department**
**Tomsk State University of Control Systems and Radioelectronics**

# ACADEMIC MOBILITY AS AN ORGANIZATIONAL MECHANISM OF INTERCULTURAL INTERACTION

E.M. Pokrovskaya, M.Y. Raitina

**Annotation**
The trends of academic mobility of students in the context of internationalization are presented. The results of a study devoted to the students' attitudes towards international education, cultural and educational exchange and academic mobility programs are presented. It is concluded that the organization of academic mobility, as a part of intercultural environment of the university, is a necessary tool of motivation for the educational process and organizational mechanism for intercultural communication.

**Introduction.** Today, the competitiveness of education is achievable only in the international context [1]. One of the important and most relevant areas of international activity is the organization of the academic mobility system as a part of the intercultural environment of the university and an organizational mechanism for intercultural communication, which is one of the most effective ways to develop educational opportunities at the individual level [2].

Currently, traveling abroad in order to gain experience in various cultural and educational practices and academic exchange is an immanent value in personal terms of young people motivated to receive education [3, 5]. The interconnection of such categories as the mobility of the individual and the social activity of the individual is obvious. Cultural and educational tourism is considered by the authors as one of the social resources of an individual aimed at improving the quality of human capital [4].

**The aim of the study.** In the presented work, attempts are made to determine the trends of academic mobility of students in the context of internationalization as part of the intercultural environment of the university, and the research methods were the method of theoretical analysis and questionnaire, through an Internet survey.

**Research results.** Structurally, learning activities can be considered as interdependent fields that include needs, motivation to learn, goals, learning activities, and self-esteem.

The organization of academic mobility, as the part of the intercultural environment of the university, acts as a necessary tool for motivating the educational process, where the leading aspects are improving the quality of education, introducing new forms and technologies of teaching, participating in the international education system, creating conditions for the subsequent expansion of intercultural interaction and the areas of introducing cultural - educational potential.

In order to reveal the opinion of the students themselves about the attitude towards international education, cultural and educational exchange and academic mobility programs, a method that is relevant for sociological projects was applied - conducting a survey via the Internet using a random sample method.

The study, conducted in January 2020 on the basis of the Tomsk State University of Control Systems and Radioelectronics (TUSUR), was attended by 298 respondents, from among students of 1-4 undergraduate courses.

The respondents were asked questions identifying the reasons for choosing to receive a foreign education and problems limiting travel; awareness of existing programs of cultural and educational exchanges.

The first question was formulated as follows: "Would you like to study abroad?" 68% of the respondents answered that it is undoubtedly "yes", because this is a prestigious way of obtaining a diploma, 22% expressed the opinion that they are patriots of their country and for 10% of respondents noted that it does not matter where to get a diploma, the main thing is its availability (Fig. 1).

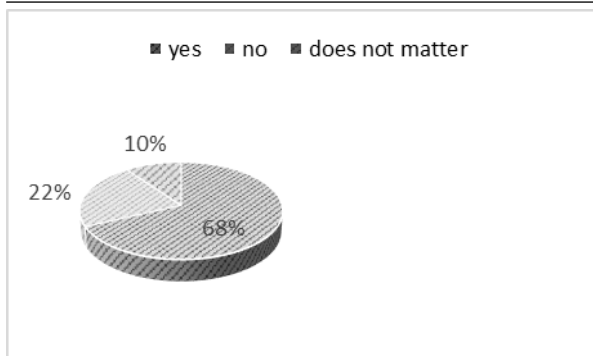

Fig. 1. Students' attitude to education abroad

The next question for respondents is: "Which country would you prefer to study?" 48% answered - "any country in Europe"; 14% - USA; 29% - Great Britain; 1% - China; 8% - did not choose any country (Fig. 2).

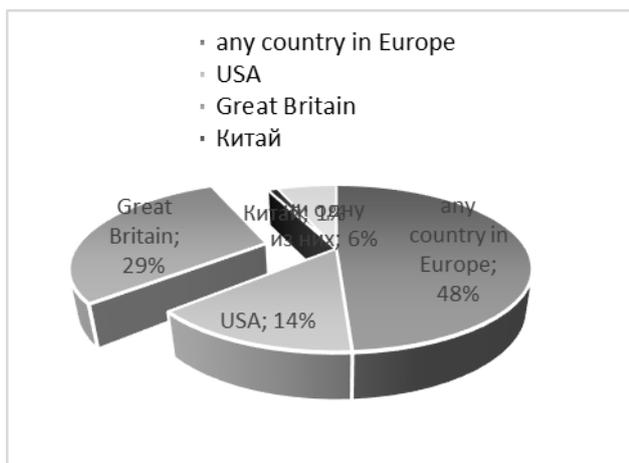

Fig. 2. Preferable country for getting education

Travel and tourism, learning languages, meeting new interesting people - these are the values of young people and factors that contribute to successful educational activities.

The next question was about the meaning of international exchanges, where 59% of respondents' answers are that international exchanges provide an opportunity to get acquainted with the culture of the country, with the local population, customs; 35% of the respondents believe that the goal of international exchanges is to obtain a prestigious education; 6% - to get a good income.

An important aspect is also the attitude towards students who came to Russia for exchange. 70% of the respondents answered with a positive attitude, they would love to learn their language, learn other traditions - cultural exchange; 19% are not interested in this issue, in other words, they are not concerned with the fact that students from other countries study in our country; but there were also those who did not like this fact - 11% have a negative attitude towards such students.

Lack of knowledge about the availability of appropriate programs that open up educational opportunities for young people is also one of the main reasons for reducing the level of exchanges. 64% of the respondents know about the existence of the "Work and Travel" program, which is popular among young people; 22% of respondents have absolutely no idea about any program, which is an indicator of the individual's low interest in learning activities; 6% are aware of the existence of the "Erasmus Mundus" program; 4% are aware of the "Global Ugrad" program; 4% know about the "IAESTE" program (Fig. 3).

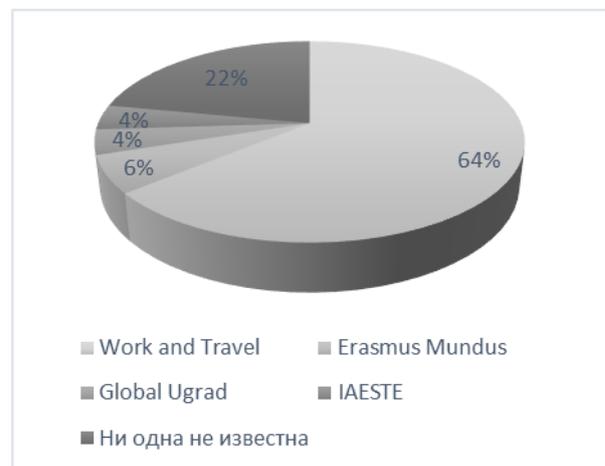

Fig. 3. Students' awareness of exchange programs

**Conclusions.** The paper identifies trends in academic mobility of students in the context of internationalization:
• the prestige of academic mobility as part of the intercultural environment of the university;
• demand for international educational programs;
• development of network international interaction;
• cultural diversity.

According to the results of this study, it is more important for respondents to get an education than financial gain (big earnings), since at the present stage in society, professional knowledge, skills, abilities, and competencies are of value, which in the future will allow them to realize their own professional and career aspirations. It is concluded that the organization of academic mobility, as a part of the intercultural environment of the university, is a necessary tool of motivation for the educational process and organizational mechanism for intercultural communication.

______________________________________

____
______________________________________

**Elena M. Pokrovskaya**
PhD in Philosophy, associate professor, head of the foreign languages department
Tomsk State University of Control Systems and Radioelectronics (TUSUR)
40, Lenina st., Tomsk, Russia, 634045
ORCID 0000-0001-9314-0077
Phone: +7 (382-2) 70-15-21
Эл. почта: pemod@yandex.ru

**Margarita Y. Raitina**
PhD in Philosophy, associate professor of the philosophy and sociology department
Tomsk State University of Control Systems and Radioelectronics (TUSUR)
40, Lenina st., Tomsk, Russia, 634045
ORCID: 0000-0002-2381-3202
Phone: +7 (382-2) 70-15-90
Email:: raitina@mail.ru